\documentclass[12pt]{iopart}
\usepackage{cite}
\usepackage{graphicx}% Include figure files
\usepackage{dcolumn}% Align table columns on decimal point
\usepackage{bm}% bold math

\begin{document}
\title[Painted loading: a toolkit for loading spatially large optical tweezer arrays]{Painted loading: a toolkit for loading spatially large optical tweezer arrays}

\author{Mitchell J Walker$^{1}$, Ryuji Moriya$^{1}$, Jack D Segal$^{1}$, Liam A P Gallagher$^{1}$, Matthew Hill$^{1}$, Frédéric Leroux$^{1}$, Zhongxiao Xu$^{1,2,3}$, Matthew P A Jones$^{1}$\footnote{Corresponding author}}

\address{$^{1}$Department of Physics, Durham University, South Road, Durham DH1 3LE, United Kingdom}
\address{$^{2}$State Key Laboratory of Quantum Optics and Quantum Optics Devices, Institute of Opto-electronics,
Shanxi University, Taiyuan, Shanxi 030006, China}
\address{$^{3}$Collaborative Innovation Center of Extreme Optics, Shanxi University, Taiyuan, Shanxi 030006, China}

\ead{m.p.a.jones@durham.ac.uk}

\date{\today}% It is always \today, today,

\begin{abstract}
Arrays of neutral atoms in optical tweezers are widely used in quantum simulation and computation, and precision frequency metrology. The capabilities of these arrays are enhanced by maximising the number of available sites. Here we increase the spatial extent of a two-dimensional array of $^{88}$Sr atoms by sweeping the frequency of the cooling light to move the atomic reservoir across the array. We load arrays with vertical heights of $>100$ \textmu{}m, exceeding the height of an array loaded from a static reservoir by a factor of $>3$. We investigate the site-to-site atom number distribution, tweezer lifetime, and temperature, achieving an average temperature across the array of $1.49(3)$~\textmu{}K. By controlling the frequency sweep we show it is possible to control the distribution of atoms across the array, including uniform and non-uniformly loaded arrays, and arrays with selectively loaded regions. We explain our results using a rate equation model which is in good qualitative agreement with the data.
\end{abstract}

\noindent{\it Keywords\/}: optical tweezer array, atomic ensembles, ultracold atoms, magneto-optical trap, atomic clock, painted loading

\submitto{Quant. Sci. Tech.}

\maketitle

\section{Introduction}
\label{sec:Introduction}

Arrays of neutral atoms in optical tweezers are a versatile platform for quantum science, enabling studies of frequency metrology~\cite{Borregaard2013,Young2020,Shaw2024}, quantum simulation~\cite{Labuhn2016,Kaufman2021} and quantum computation~\cite{Weiss2017,Henriet2020,Wintersperger2023}. For these applications, key figures of merit - such as available information storage, for quantum computation~\cite{Henriet2020}, or statistical errors, for frequency metrology~\cite{Wineland1994} - scale favourably with increasing atom number. Therefore, an ongoing goal is to generate and load increasingly large arrays, with the state of the art in alkali atoms exceeding 1000 sites~\cite{Manetsch2024,Pause2024,Pichard2024}.

In multivalent atoms, such as Sr~\cite{Cooper2018,Norcia2018,Covey2019,Norcia2019,Jackson2020,Urech2022}, Yb~\cite{Saskin2019,Reichardt2024,Nakamura2024}, Er~\cite{Grun2024}, and Dy~\cite{Bloch2023,Biagioni2025}, electronic transitions between singlet and triplet spin states give rise to narrow spectral features ($\Gamma< 1$~Hz). These transitions are ideal for frequency metrology~\cite{Derevianko2011}, having already allowed for fractional uncertainty limits in Cs atomic clocks to be surpassed by several orders of magnitude~\cite{Aeppli2024}. In addition, high-lying Rydberg states can be used to entangle atoms in tweezer arrays, enabling measurements beyond the standard quantum limit~\cite{Gil2014,Eckner2023,Bornet2023,Hines2023,Finkelstein2024}. These features also make arrays using these species an excellent platform for quantum computing and simulation. Qubits based on the singlet-triplet transitions offer high-fidelity readout and error correction~\cite{Madjarov2020,Wu2022,Ma2022,Nakamura2024}, while entanglement using the Rydberg states allows multi-qubit gate operations to be performed~\cite{Evered2023,Cao2024,Xu2024}.

The number of tweezer array sites that can be loaded without re-arrangement is determined by the dimensionality and spacing of the array, and its overlap with the cold-atom reservoir used for loading (figure~\ref{Fig:Pictograph}). In multivalent atoms, this reservoir is typically a magneto-optical trap (MOT) operating on a narrow intercombination transition, in which atoms can be cooled to $<1$~\textmu{}K~\cite{Loftus2004}. Narrow-line MOTs (nMOTs) can be characterised by the parameter $\eta=\Gamma/\omega_\mathrm{R}$ where $\Gamma$ is the natural linewidth of the cooling transition and $\omega_\mathrm{R}$ is the frequency shift due to atomic recoil following the absorption or emission of a photon~\cite{Hanley2017}. In nMOTs, $\eta\approx1$, and the recoil from an individual scattering event significantly alters the probability of subsequent absorption. As a result, nMOTs have a characteristic elliptical shape, where gravity causes the atoms to be confined to a shell around the resonance condition~\cite{Loftus2004,Hanley2017}. In strontium, where $\eta = 1.6$ for nMOTs operating on the $^1S_0\rightarrow{}^3P_1$ transition, this is particularly apparent, with atoms in the nMOT confined vertically to a shell of roughly 10~\textmu{}m thickness~\cite{Hanley2017}.

The reduced vertical spatial extent of nMOTs limits the size of tweezer arrays and lattices which can be loaded, with arrays typically consisting of order 100 sites~\cite{Jenkins2022,Biagioni2025,Cooper2018}. To mitigate this, one solution is to restrict the tweezer array to the horizontal plane, where the spatial extent of the MOT is larger~\cite{Norcia2018,Cooper2018,Holman2024}. However, this is limited to quasi-two-dimensional arrays, and imposes restrictions on the experiment's geometry. The spatial extent of the nMOT can be increased by modulation of the cooling laser, artificially broadening the transition~\cite{Norcia2018,Snigirev2019}, although this also increases temperature. Elsewhere, optical transport via re-arrangeable arrays has been used to move atoms beyond the nMOT's spatial extent~\cite{Barnes2022,Knottnerus2025,Yan2025}. Large ($>1$~mm) optical lattices have been loaded with Yb atoms by dropping and recapturing the MOT\cite{Hassan2024}. However, this method cannot be implemented using Sr atoms due to the narrower linewidth of the cooling transition.

In this work, we present an alternative method to load large ($>100$~\textmu{}m) $^{88}$Sr tweezer arrays. By sweeping the detuning of the cooling light, we move the reservoir over the tweezer array during loading as shown in figure~\ref{Fig:Pictograph}. By controlling the speed and shape of the sweep, we demonstrate control over the site-to-site distribution of atoms in the array. This includes uniform and non-uniformly loaded arrays, as well as selective loading of individual regions. Our method builds on related work on an optical lattice loaded using linear sweeps of fixed speed \cite{HobsonThesis}. We present a rate equation model which is found to be in good qualitative agreement with the data, and shows that the underlying physics is governed by the interplay of ramp speed and trap loss arising from heating and cooling mechanisms in the tweezer and nMOT.

\section{Methods}
\label{Sec:Methods}

\begin{figure}
    \includegraphics[width=\linewidth{}]{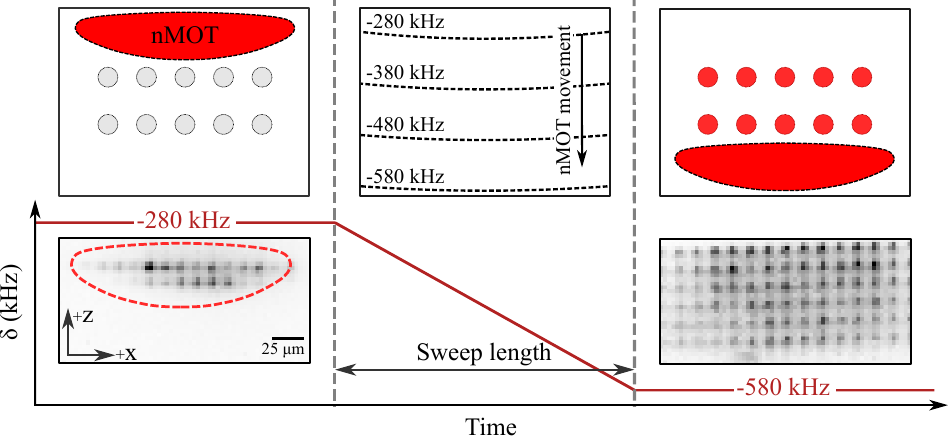}
    \caption{Pictographic representation of painted loading. The nMOT is initially placed above the tweezer array (top left). This reservoir is then swept across the array by changing the frequency of the cooling light, moving the nMOT resonance condition across the array (top middle); tweezer sites crossed by the nMOT are loaded with atoms (top right). For illustrative purposes, the field of view of the cartoon is larger than that of our imaging system. \textbf{Bottom-left inset:} Image of a 90-site tweezer array loaded without sweeping. Only sites within the confines of the nMOT can be loaded. \textbf{Bottom-right inset:} The same array loaded using painted loading, showing atoms in all of the tweezer sites.}
    \label{Fig:Pictograph}
\end{figure}

We study the loading of an arbitrary 2D array of optical tweezers from an nMOT of $^{88}$Sr atoms. Details of the experimental setup and the steps taken to load the nMOT are described in \cite{HanleyThesis,HillThesis}.

We define the $x(y)$ axes as perpendicular(parallel) to the tweezer beam axis in the horizontal plane, with $-z$ the direction of gravity. The atoms in the nMOT are confined to a bowl-shaped shell around the resonance condition with measured FWHMs of $\Delta{}x_\mathrm{FWHM}=171(8)$~\textmu{}m and $\Delta{}z_\mathrm{FWHM}=29(2)$~\textmu{}m along the $x$ and $z$ axes respectively. The nMOT is generated using a quadrupole magnetic field with a gradient in the vertical direction of $B^\prime_\mathrm{z}=8.9(2)$~G~${\mathrm{cm}}^{-1}$ and three retroreflected beams red-detuned from the $\mathrm{5s}^2\,^1S_0\rightarrow{}\mathrm{5s5p}\,^3P_1$ transition at 689~nm. The total nMOT beam power is split such that $60\%$ is in the vertical beam pair and $20\%$ is in each of the horizontal beam pairs, giving an atom number of approximately $10^5$. For the results presented in this work, we use a combined power in the nMOT beams of $64$~\textmu{}W and a $1/e$ beam waist of $0.255(15)$~cm, resulting in an atom temperature of $1.99(8)$~\textmu{}K and an nMOT lifetime of $97.0(6)$~ms. These parameters are dependent on the intensity of the nMOT beams. By reducing the combined power to $16$~\textmu{}W, we obtain a significantly reduced temperature of $0.65(3)$~\textmu{}K, at the expense of reducing the lifetime to 19.00(15)~ms. 

By controlling the detuning $\delta$ of the nMOT beams, it is possible to vary the vertical position $z$ of the resonance condition ($h \delta = g_\mathrm{J} \mu_\mathrm{B} B^\prime_\mathrm{z} z$, where $g_\mathrm{J}=3/2$ is the relevant Land\'{e} factor and $\mu_\mathrm{B}$ is the Bohr magneton) from the magnetic field zero point and thus move the reservoir of atoms (figure~\ref{Fig:Pictograph}). The position of the nMOT resonance condition varies linearly with detuning over the range of frequencies considered in this work~\cite{Hanley2017}; we measure the change in vertical position with respect to frequency as $0.53(1)$~\textmu{}m~kHz$^{-1}$, which is in agreement with the expected value.

We generate our tweezer arrays at a magic wavelength for the 698~nm clock transition ($813.427$~nm~\cite{Katori2009}) using a liquid-crystal spatial light modulator (SLM) (Hamamatsu LCOS-SLM X10468-02). The procedure for designing the SLM phase mask and equalising the trap depths is described in \cite{Nogrette2014}.
The tweezers are focused to a waist of $2.21(3)$~\textmu{}m at a distance of $37$~mm from an aspheric in-vacuo focusing lens (numerical aperture $\mathrm{NA}=0.26(2)$ \cite{Jackson2020}). We load ensembles of atoms into the tweezers at a trap depth of $U_{0}/k_\mathrm{B}=9.9(2)$~\textmu{}K before ramping to a final depth of $U_\mathrm{F}/k_\mathrm{B}=145(3)$~\textmu{}K over $100$~ms for imaging. We measure a background-gas-limited lifetime of $7.4(3)$~s for atoms in $10$~\textmu{}K tweezers.

To load our tweezers, we begin with the nMOT positioned at approximately $z=+10$~\textmu{}m above uppermost site of the tweezer array (figure~\ref{Fig:Pictograph}). By increasing the detuning of the cooling light we sweep the atoms in the nMOT over the tweezer sites, moving them in the $-z$ direction. Once loading is complete, the nMOT beams are extinguished and any remaining unloaded atoms are allowed to fall away. We liken the nMOT to a paint roller passing over the array and filling in sites, and thus name the method ``painted loading''. As outlined below, the final distribution of atoms across sites in the array depends on the speed at and distance over which the atoms are moved. The detuning of the nMOT beams is controlled via a double-pass acousto-optical modulator.

The loaded array is fluorescence imaged using a probe beam acting on the $5\mathrm{s}^{2\:1}S_{0}\rightarrow 5\mathrm{s}5\mathrm{p}^{\:1}P_{1}$ transition. The in-vacuo lens which focuses the tweezer array also acts as our collection optic, and the probe beam polarisation is oriented to maximise the fraction of fluorescence collected by this lens. The collected photons are imaged using a single-photon avalanche diode array camera (Micro Photon Devices PC-2D-32X64) \cite{Jackson2020}, and our imaging system has an overall detection efficiency of $0.8(1)\%$.

\section{Loading Tweezer Arrays}
\label{sec:Loading}
Figure~\ref{Fig:Pictograph} demonstrates the utility of the painted loading technique. Here we show a 90-site rectangular tweezer array first loaded from a static nMOT (bottom left). Only two of the rows in the array are loaded due to the limited spatial extent of the nMOT. When the painted loading technique is used (bottom right) it is possible to load the entire 90-site array, filling the 200~\textmu{}m by 100~\textmu{}m field of view of our imaging system.

\subsection{Varying sweep speed}
\label{subsec:LinearSweeping}

\begin{figure}
    \centering
    \includegraphics[width=0.95\linewidth{}]{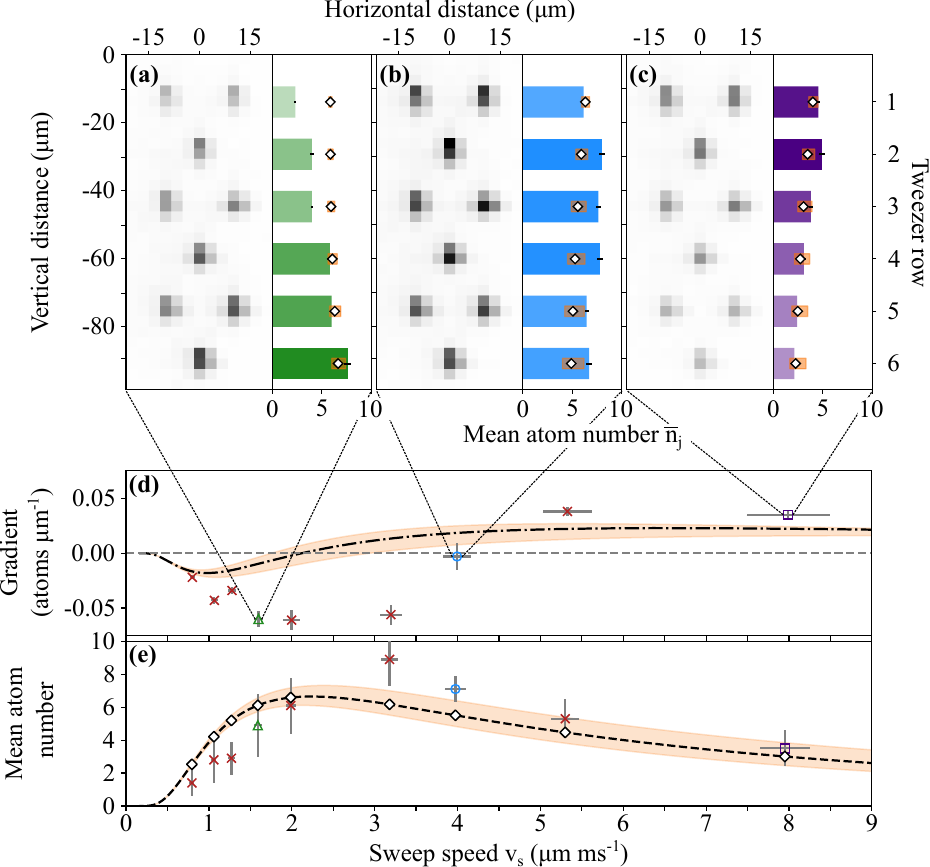}
    \caption{\textbf{(a, b, c)}~Averaged fluorescence images showing atom distributions loaded at sweep speeds of (a) $1.59(3)$~\textmu{}m~ms$^{-1}$ (b) $3.97(8)$~\textmu{}m~ms$^{-1}$ and (c) $7.95(15)$~\textmu{}m~ms$^{-1}$. Bar charts show the corresponding $\bar{n}_j$ by row, with model results (uncertainties) marked as diamonds (orange shaded regions). Each array was imaged 30 times. \textbf{(d)}~Measured gradient of $\bar{n}_j$ as a function of sweep speed. Grey dashed line denotes zero gradient (uniform loading). Black dashed line (orange shaded area) shows the model predicted gradient (uncertainty) as a function of sweep speed. \textbf{(e)}~Array-averaged number of atoms per site as a function of sweep speed. Black dashed line shows the model best fit at a scaling factor of $\kappa N_\mathrm{M}(0) =5.7(6)\times10^{-9}$, with orange shaded area showing the uncertainty.}
    \label{Fig:Atom Distributions}
\end{figure}

By varying the speed at which the atoms are swept across the tweezer array, it is possible to obtain row-by-row control of the number in each site of an array (figure~\ref{Fig:Atom Distributions}). Panels (a)-(c) show a 9-site array loaded at three different sweeps speeds with bar charts showing the mean atom number per site $\bar{n}_j$ in each row. Diamonds show predictions from our rate equation model (see Section \ref{sec:model}). At low sweep speeds (figure~\ref{Fig:Atom Distributions}(a)) more atoms are loaded into sites lower in the array (loaded later), while for fast sweeps (figure~\ref{Fig:Atom Distributions}(c)) we find the opposite. At an intermediate sweep speed (figure~\ref{Fig:Atom Distributions}(b)), we find the atom number distribution is roughly uniform across the array. 

To quantify the distribution of atom number across the array, we use a linear fit to the atom number per row to extract the gradient in atom number as a function of vertical position. Figure~\ref{Fig:Atom Distributions}(d) shows how this gradient varies as a function of sweep speed $v_\mathrm{s}$, highlighting the change from a negative gradient (more atoms in sites loaded later) to a positive gradient (more atoms in sites loaded earlier) as the sweep speed increases. We also find that the average number of atoms per site loaded into the array varies significantly with sweep speed, as shown in figure~\ref{Fig:Atom Distributions}(e). Here, we find that the number of atoms loaded into the array is maximum for intermediate sweep speeds, close to the speeds where the array is uniformly loaded. We attribute the variation in atom number distributions to the loading and loss rates into the tweezers which vary as the nMOT is swept over the sites. In the following, we experimentally measure parameters relating to these rates before constructing a model based on rate equations.

\subsection{Lifetime}

\begin{figure}
    \centering
    \includegraphics[width=0.95\linewidth{}]{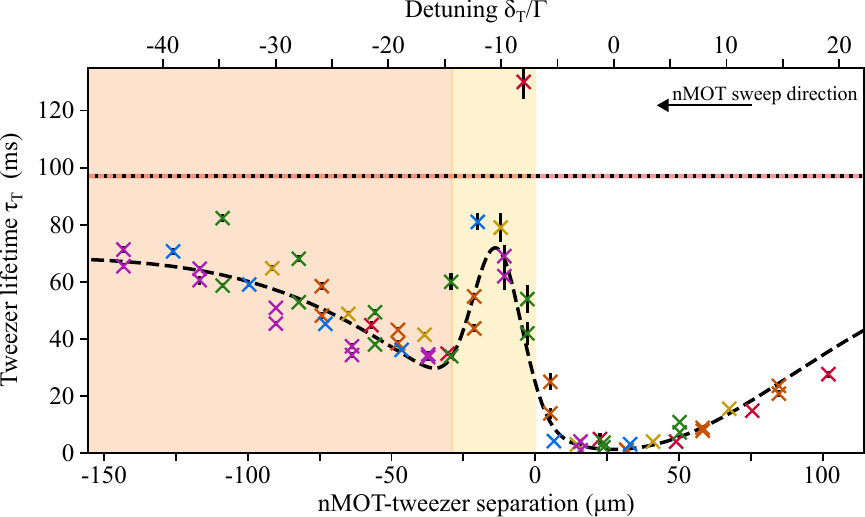}
    \caption{Measured tweezer lifetime $\tau_\mathrm{T}$ as a function of $\delta_\mathrm{T}$ and the corresponding nMOT-tweezer separation. Different colour crosses are used to differentiate data points for tweezers at different $z$ positions. The yellow and orange shaded regions, corresponding to regions 2 and 3 in equation~\ref{Eq:Regions} respectively, are relevant for painted loading. A dotted black line indicates the measured nMOT lifetime for this data set, $97.0(4)$~ms, with shading denoting the uncertainty. The dashed black line shows the line of best fit for equation~\ref{Eq:Tweezer Detuning} to the data.}
    \label{Fig:Lifetimes}
\end{figure}

To gain insight into the rate of loss of atoms loaded into our tweezers during painted loading, we measured the trap lifetime in the tweezers $\tau_\mathrm{T}$ in the presence of cooling light at different detunings. The 9-site tweezer array was loaded with the linear sweep shown in figure \ref{Fig:Atom Distributions}(b), and then the cooling light was temporarily extinguished to allow remaining atoms in the nMOT to fall away. The cooling light was then switched back on at detuning $\delta$ and the trap lifetime in each site was measured. Since the quadrupole field remained switched on, atoms in a site at position $z_{\mathrm{T}}$ experienced a detuning given by $\delta_\mathrm{T} = \delta + \Delta_\mathrm{ls} - \left(g_\mathrm{J} \mu_\mathrm{B} B_\mathrm{z}^\prime z_{\mathrm{T}}\right)/h$, where $\Delta_\mathrm{ls}=-49$~kHz is the differential lightshift due to the tweezer light at a trap depth of $U_{0}/k_\mathrm{B}=10$~\textmu{}K~\cite{Urech2022}.

The results are shown in figure~\ref{Fig:Lifetimes}. We find that the tweezer lifetime varies significantly as a function of nMOT beam detuning, whereas the nMOT lifetime remains constant over this range of frequencies. When plotted against $\delta_\mathrm{T}$, the measured tweezer lifetimes from different rows collapse onto a single curve consisting of two features: a broad reduction in trap lifetime centred on $\delta_\mathrm{T}=0$ that we attribute to heating from spontaneous emission, and a narrow region of enhanced lifetime centred at $\delta{}\approx{} -10\Gamma{}$ that we attribute to the attractive Sisyphus cooling mechanism previously studied in magic-wavelength Sr tweezer arrays \cite{Ivanov2011,Covey2019, Urech2022}. We note that the lifetime falls towards zero at $\delta_\mathrm{T}=0$, which means the painted loading technique only works if the detuning is swept away from resonance (nMOT moves downwards). 

\subsection{Temperature}

\begin{figure}
    \centering
    \includegraphics[width=0.95\linewidth{}]{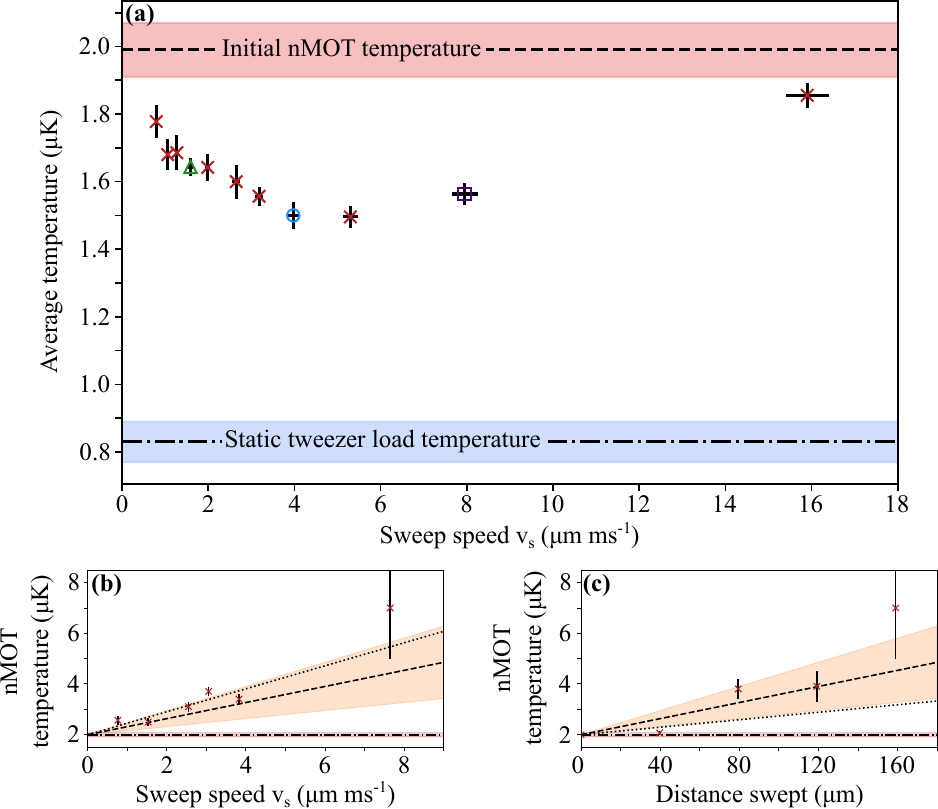}
    \caption{\textbf{(a)}~Mean temperature of atoms in the array as a function of sweep speed $v_\mathrm{s}$. Points corresponding to figures~\ref{Fig:Atom Distributions}(a), (b), (c) are highlighted. The initial nMOT temperature, $T_\mathrm{M}(0)=1.99(8)$~\textmu{}K, and the temperature achieved for loading from a static nMOT, $0.83(6)$~\textmu{}K, are marked as dashed and dash-dotted lines respectively, with the corresponding uncertainties represented by shading. \textbf{(b, c)}~Final nMOT temperature as a function of (b) sweep speed for a fixed sweep distance of $159(3)$~\textmu{}m and (c) distance swept for a fixed sweep speed of $7.95(15)$~\textmu{}m~ms$^{-1}$. $T_\mathrm{M}(0)$ is indicated as a dash-dotted line. Dotted lines show linear fits to the data. Dashed lines indicate the nMOT heating rate used in the model (see Section~\ref{sec:model}), with orange shaded region showing the uncertainty.}
    \label{Fig:Temperature}
\end{figure}

Measured average temperatures for atoms loaded into the tweezer array at different sweep speeds are shown in figure~\ref{Fig:Temperature}. Temperatures were measured via the release and recapture method~\cite{Tuchendler2008}. We observe a clear variation with sweep speed. At both extremes, the temperature approaches that of the initial nMOT, while a minimum occurs for intermediate speeds that is approximately twice as high as that obtained from static loading. We measure an average normalised standard deviation in atom temperatures across the array of $7.3(5)$\%, but do not find a systematic positional dependence for any sweep speed. 

We explain the variation of temperature with sweep speed as follows: when the atoms are initially loaded into the tweezer from the nMOT, they are cooled by the nMOT beams via an attractive Sisyphus mechanism, towards a minimum temperature set by the Sisyphus cap~\cite{Covey2019,Urech2022}. This Sisyphus cooling occurs over a narrow range of red-detuned frequencies, outside of which the nMOT beams instead act to heat the atoms back out of the trap. At fast sweep speeds, the detuning changes too quickly to significantly cool the atoms, and at slow sweep speeds, the atoms conversely spend a long time being heated after the initial cooling period. We attribute the colder final temperature  achieved via static loading to the atoms being cooled for a longer period of time ($20$ ms) than in the painted loading case, and due to the absence of heating from the nMOT beams after loading.

As shown in figures~\ref{Fig:Temperature}(b) and (c), the effect of the detuning sweep on the temperature of the nMOT reservoir is much more significant, with strong heating for all but the slowest sweeps. A comparison of figures~\ref{Fig:Temperature}(a) and (b), (c) emphasises the importance of Sisyphyus cooling mechanisms in the tweezers, as the final temperature of atoms in the tweezers is significantly lower than that of those in the nMOT. We discuss the impact of the linear dependence of final nMOT temperature on sweep speed in Section~\ref{sec:model}.

\subsection{Non-linear sweeping}
\label{subsec:Non-linear}

In tweezer array experiments, it is preferable to minimise the duration of the loading phase to maximise the experimental duty cycle. Here we extend our technique to non-linear sweeps where $v_\mathrm{s}$ varies during the sweep. In this way, we reduce the loading time of uniform and linear gradient arrays to 20 ms and enable other novel atom distributions.

\begin{figure}
    \centering
    \includegraphics[width=0.95\linewidth{}]{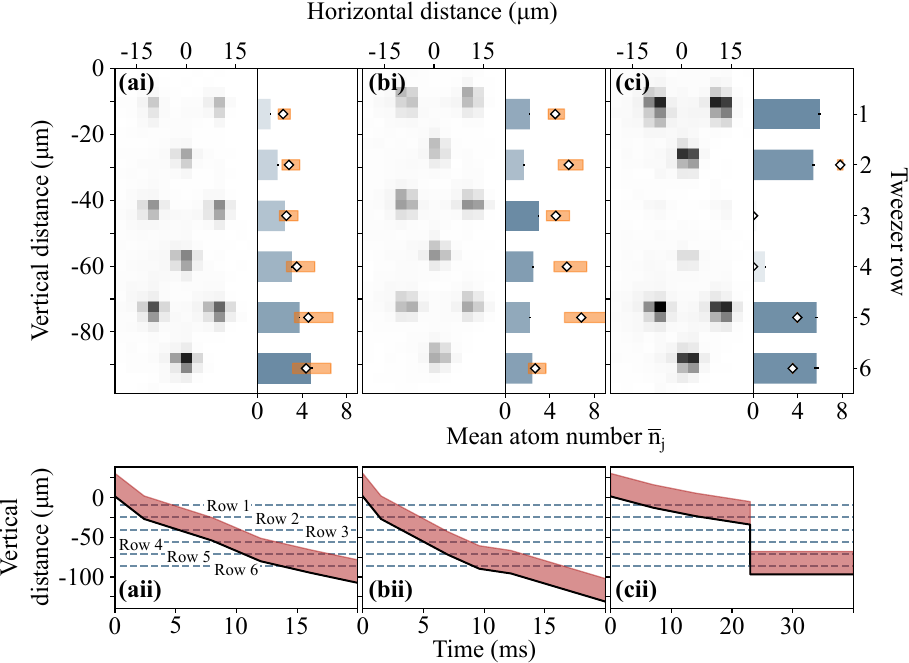}
    \caption{\textbf{(i)} Averaged fluorescence images for arbitrary sweeps targetting (a) linearly varying, (b) uniform, and (c) stepped atom distributions. Bar charts show the corresponding $\bar{n}_j$ by row, with model results marked as diamonds. The model prediction for the number of atoms in row 1 of (c) was $19.8(5)$ atoms, and as such is not displayed on these axes. Each array was imaged 250 times. \textbf{(ii)}~Corresponding variation in the position of the nMOT resonance condition (black solid line) as a function of time. Red shading indicates the approximate region covered by the nMOT, and dashed blue lines indicate the position of each row in the tweezer array.}
    \label{Fig:Non-Uniform Loading}
\end{figure}

Figure~\ref{Fig:Non-Uniform Loading}(a) shows a linear gradient in atom number across the array loaded using a 20~ms non-linear sweep. We are able to achieve a gradient in atom number across the array of -0.043(1)~atoms~\textmu{}m$^{-1}$, which is slightly lower than that achieved in figure~\ref{Fig:Atom Distributions}(a) (-0.060(7)~atoms~\textmu{}m$^{-1}$), but with a reduction in sweep duration from 100~ms to 20~ms. We note that the reduction in gradient is mostly dominated by the atom number loaded, which has reduced from an average of $7.6(4)$ atoms in the bottom site to $4.8(2)$ atoms. 

Figure~\ref{Fig:Non-Uniform Loading}(b) shows a uniformly loaded tweezer array in a 20~ms sweep time. The variance of the mean atom number across the array is 7.6\%, compared to 11.0\% for figure~\ref{Fig:Atom Distributions}(b). The sweep time has been reduced by $50\%$ from $40$~ms to $20$~ms. However, we note that the mean atom number in the array has dropped by a factor of about 3: we attribute this to a reduced initial number of atoms in the nMOT, due to experimental fluctuations.

It is possible to design a frequency sweep to selectively load only certain sites, as shown in figure~\ref{Fig:Non-Uniform Loading}(c). Here, we include a step change in frequency, causing the resonance condition of the nMOT to jump. The atoms in the nMOT then fall and are recaptured at the new resonance condition. We find that while the atoms are in freefall, there is no significant scattering of light from the nMOT beams by the atoms. As such, tweezer sites located in the region over which the atoms fall are loaded with far fewer atoms than other sites in the array. For the array shown in figure~\ref{Fig:Non-Uniform Loading}(c), sites in rows which the nMOT is stepped over (3, 4) have an overall average occupation of $12.1(4)$\% of that of sites in other rows (1, 2, 5, 6).

The non-linear sweeps shown in figures~\ref{Fig:Non-Uniform Loading}(d)-(f) were assembled from short linear sections and optimised manually. We envisage that tighter control over the atom number distribution and faster loading could be achieved by applying feedback and optimisation algorithms to iteratively optimise the shape and duration of the sweep to match an arbitrary target distribution. This is beyond the scope of the work presented here; we note that machine learning has already been successfully applied to similar problems in cold-atom experiments \cite{Wigley2016}.

\section{Rate equation model}\label{sec:model}
To give further insight into the underlying physics of painted loading, we have developed a model based on rate equations. We consider three distinct time regions: region 1, before the tweezer is loaded; region 2, during loading; and region 3, after the tweezer is loaded. We define $t=0$ as when the nMOT begins sweeping, $t=t_{1,j}$ and $t=t_{2,j}$ as the times at which a tweezer in row $j$ begins and ends loading respectively (when the nMOT-tweezer separation is $0$ and $-\Delta{}z_\mathrm{FWHM}$), and $t=t_{\mathrm{s}}$ as the total sweep time. The rate of change of atom number in a tweezer in row $j$ is

\begin{equation}
\left(\frac{dN_{\mathrm{T}}}{dt}\right)_{j}=\cases{0 & Region 1, $\:\: 0 \le{} t < t_{1,j} $\\
        R_\mathrm{load}-R_\mathrm{loss} & Region 2, $\:\:t_{1,j} \le{} t \le{} t_{2,j}$ \\
        -R_\mathrm{loss} & Region 3, $\:\: t_{2,j} < t \le{} t_{\mathrm{s}}$\\ }
        \label{Eq:Regions}
\end{equation}

\noindent{}where $R_{\mathrm{load}}$ and $R_{\mathrm{loss}}$ are the tweezer loading and loss rates respectively.

We model the loading rate as

\begin{equation}            
    R_{\mathrm{load}}\left(t\right)=\kappa{}\left[\frac{N_\mathrm{M}\left(t\right)}{{T_\mathrm{M}(t)}^{3/2}}\right],
    \label{Eq:Rload}
\end{equation}

\noindent{}where $N_\mathrm{M}$ and $T_\mathrm{M}$ are the number of atoms in and temperature of the nMOT respectively and $\kappa{}$ is a constant of proportionality. The dependence of the loading rate on the temperature used in our model is based on simulations presented in~\cite{Watson2021}. For our experimental parameters, $\kappa{}<<1$, and as such the number of atoms lost from the nMOT by loading the tweezer is negligible.

For a MOT with initial atom number $N_{\mathrm{M}}(0)$, the number of atoms in the MOT at time $t$ is given by

\begin{equation}
N_{\mathrm{M}}\left(t\right) = N_{\mathrm{M}}(0)\mathrm{exp}\left(-\frac{t}{\tau{}_{\mathrm{M}}}\right),
\end{equation}

\noindent{}where $\tau{}_\mathrm{M}$ is the lifetime of the atoms in the nMOT, which we have confirmed experimentally to be constant during painted loading.

As the nMOT is swept over the array, its temperature $T_\mathrm{M}$ increases. We model the temperature increase during the sweep as being linear in both distance swept $\Delta z$ and sweep speed $v_\mathrm{s}$, in agreement with the data shown in figures~\ref{Fig:Temperature}(b) and (c). As $\Delta z = v_\mathrm{s} t$, the temperature varies quadratically during the sweep

\begin{equation}
    T_\mathrm{M}(t) = \beta_{\mathrm{M}}\,{v_{\mathrm{s}}}^{2}t + T_{\mathrm{M}}(0),
\end{equation}

\noindent{}where $\beta_{\mathrm{M}}=2(1)\times{}10^{-3}$~\textmu{}K~(\textmu{}m~ms$^{-1}$)$^{-2}$~ms$^{-1}$ is obtained from linear fits to the data in figures~\ref{Fig:Temperature}(b) and (c), and $T_{\mathrm{M}}(0)=1.99(8)$~\textmu{}K is measured using ballistic expansion. We take the uncertainty on $\beta{}_\mathrm{M}$ to be the dominant source of uncertainty in our model parameters, and so use this to place uncertainty bounds on our model-predicted (orange shaded regions in figures).

For results presented in figure~\ref{Fig:Non-Uniform Loading}(c), it is necessary to consider how the atoms in the nMOT heat when the frequency is stepped discontinuously from one value to another. For these frequency changes, the atoms no longer move with the resonance condition but instead fall under gravity. Here, we say that the temperature of the atoms increases by an amount

\begin{equation}
    \Delta{}T_{\mathrm{M}}=\frac{2 m g \Delta z}{3k_{B}},
\end{equation}

\noindent{}once the atoms are recaptured by and thermalise in the nMOT, where $m$ is the mass of an atom, $\sqrt{2 g \Delta z}$ is the change in speed of the atoms over the drop, and $k_{B}$ is the Boltzmann constant. For drops larger than $\Delta z \sim 20$~\textmu{}m, the atoms accelerate to speeds where they can no longer be recaptured by the nMOT potential and are lost.

$R_{\mathrm{loss}}$ depends on the number $N_{\mathrm{T}}$ and lifetime $\tau{}_{\mathrm{T}}$ of atoms in a tweezer, and is modelled as
\begin{equation}
    R_\mathrm{loss}\left(t\right)=N_{\mathrm{T}}\left(t\right)\mathrm{exp}\left[-\frac{t}{\tau{}_{\mathrm{T}}\left(\delta{}_{\mathrm{T}}\right)}\right].
\end{equation}

\noindent{}As shown in figure~\ref{Fig:Lifetimes}, the lifetime $\tau_\mathrm{T}(\delta_\mathrm{T})$ has a strong detuning dependence and therefore varies with time as $\delta_\mathrm{T}$ is swept. We model $\tau_\mathrm{T}(\delta_\mathrm{T})$ using the sum of two Gaussian lineshapes

\begin{equation}
    \tau{}_\mathrm{T}(\delta{}_\mathrm{T})=\tau{}_{0} - A_{\mathrm{heat}}\mathrm{exp}\left[-\frac{{\delta{}_{\mathrm{T}}}^{2}}{2{\sigma{}_{\mathrm{heat}}}^{2}}\right]+A_{\mathrm{cool}}\mathrm{exp}\left[-\frac{\left(\delta{}_{\mathrm{T}}-\delta{}_\mathrm{cool}\right)^{2}}{2{\sigma{}_{\mathrm{cool}}}^{2}}\right],
    \label{Eq:Tweezer Detuning}
\end{equation}

\noindent{}where the subscripts heat, cool refer to the broad and narrow features respectively. A least-squares fit of equation~\ref{Eq:Tweezer Detuning} is in excellent agreement with the experimental data as shown by the dashed line in figure~\ref{Fig:Lifetimes}, giving $\tau{}_{0}=69(4)$~ms, $A_\mathrm{heat}=67(4)$~ms, $\sigma{}_\mathrm{heat}=119(6)$~kHz, $A_\mathrm{cool}=58(12)$~ms, $\delta{}_\mathrm{cool}=-61(3)$~kHz, and $\sigma{}_\mathrm{cool}=16(2)$~kHz.

To predict the number of atoms loaded into a given tweezer site, we numerically integrate over the rate equations. Values for $v_{s}$, $t_{1,j}$, $t_{2,j}$, and $t_{s}$ are set by our experimental parameters. The remaining free parameter is the product $\kappa N_\mathrm{M}(0)$ that appears in the loading rate (equation~\ref{Eq:Rload}). This parameter amounts to an overall scaling of the final atom number - it does not affect the relative number of atoms in each site. As the nMOT is optically thick, we cannot reliably estimate $N_\mathrm{M}(0)$ from our fluorescence images, and therefore we determine a value of $\kappa{} N_\mathrm{M} (0) = 5.7(6)\times10^{-9}$ by fitting the model results to the average atom number data in figure~\ref{Fig:Atom Distributions}(e).

The resulting predicted atom number distributions are compared to the data for linear sweeps in figure~\ref{Fig:Atom Distributions}. The model predicts the trends in average atom number (figure~\ref{Fig:Atom Distributions}(e)) well, and correctly predicts the positive, uniform and negative gradients for different sweep speeds in figure~\ref{Fig:Atom Distributions}(a)-(c). The model qualitatively predicts the observed trend in atom distribution gradients (figure~\ref{Fig:Atom Distributions}(d)), but not quantitatively. The model is also in good qualitative agreement with the non-linear frequency sweeps presented in figure~\ref{Fig:Non-Uniform Loading}. For the results in figure \ref{Fig:Non-Uniform Loading}(b), the measured average atom number per row is $48(7)$\% that of the model-predicted value. We explain this discrepancy as being due to a similar reduction in the initial number of atoms in the nMOT for this data set (reduced to $43.8(9)$\% of the value associated with figure~\ref{Fig:Atom Distributions}, based on total average counts measured in fluorescence images of the nMOT), which was not accounted for in our model. Other discrepancies between measured and predicted values are attributed to our simplified consideration of the dynamics of atoms in the nMOT; we neglect the precise mechanics of how the atoms' temperature and density changes as the nMOT is swept, which we take to be beyond the scope of this work.

\section{Discussion}
We have shown that painted loading provides not only a way of extending the size of tweezer array that may be loaded, but also a method for controlling the atom number distribution. The technique is simple to implement, requiring only the ability to approximately equalise the lifetime of atoms in the nMOT and tweezer array (when exposed to 689 nm light), and the ability to adjust the frequency of the nMOT beams in a linear or arbitrary way during the experimental sequence. We have also shown that there is a minimal cost with respect to the temperature of atoms loaded into the array. The increased atom temperature due to painted loading could be remedied by the inclusion of subsequent Sisyphus cooling~\cite{Covey2019, Urech2022}.

For experiments with arrays of small ensembles, painted loading provides a way of controlling effects that depend on atom number or density, such as clock shifts or spin squeezing \cite{Gil2014}. A gradient across the array enables these effects to be studied systematically within a single experiment, while uniform arrays enable such effects to be more tightly controlled. Selective loading of parts of the array provides an alternative to using complex array geometries or sorting atoms for quantum computing and metrology experiments that require distinct sub-arrays~\cite{Shaw2024,Evered2023}.

The success of the rate equation model shows that the variation of the atom number distribution with sweep speed arises from two effects. The first is the heating of the atoms in the nMOT as they are swept over the array. This heating rate acts to reduce the loading rate of atoms from the nMOT into tweezer sites as the nMOT is swept. The second effect is that the lifetime of atoms in the nMOT is chosen to be longer than that of those in a tweezer site when bathed in the nMOT beams. These two effects are both necessary in order to see a sign change in the gradient of atom distributions in the array as a function of sweep speed.

\begin{figure}
    \includegraphics[width=\linewidth]{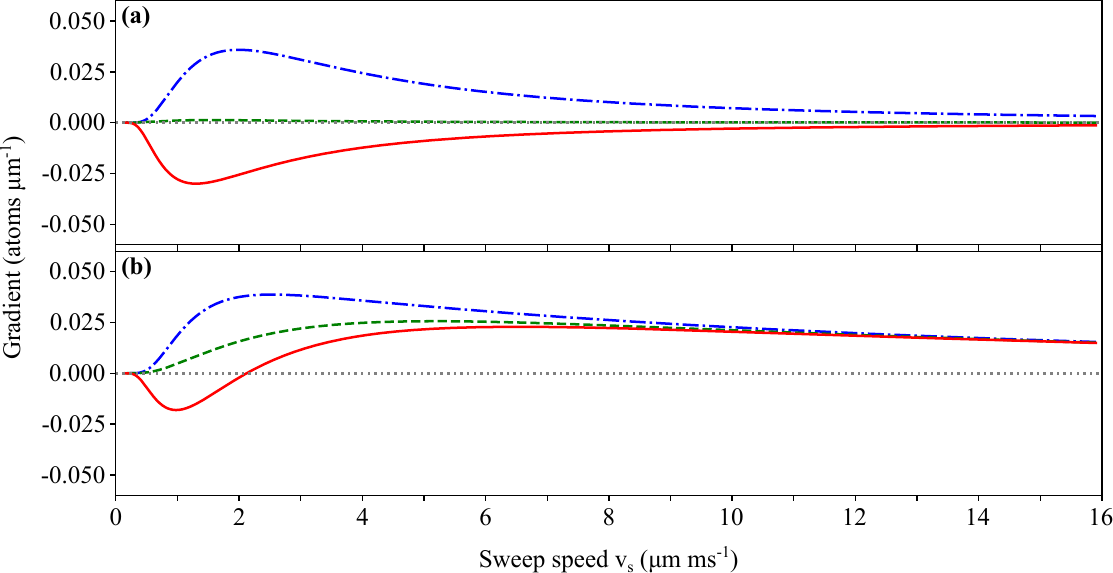}
    \caption{Model predicted gradients as a function of sweep speed for (a) no nMOT heating, i.e. $\beta{}_{\mathrm{M}}=0$ and (b) nMOT heating at a rate $\beta{}_{\mathrm{M}}=2\times{}10^{-3}$~\textmu{}K~(\textmu{}m~ms$^{-1}$)$^{-2}$~ms$^{-1}$. Predictions are shown for nMOT lifetimes of $30$~ms (blue dot-dashed line), $60$~ms (green dashed line), and $97$~ms (red solid line). Grey dashed line denotes zero gradient (uniform loading).}
    \label{Model Predictions}
\end{figure}

Figure~\ref{Model Predictions} shows model-predicted gradient curves for different nMOT lifetimes in the case that nMOT heating is (a) neglected or (b) included in the model. We highlight that only in the case that both nMOT heating is included and that the nMOT lifetime is long compared to tweezer lifetime - as is the case for the solid red line in figure~\ref{Model Predictions}(b) - is a gradient flip predicted by the model.

Experiments with single atoms use collisional effects in much smaller tweezers (waist $< 1$~\textmu{}m) to restrict the atom number at each site to either zero or one. In Sr, this typically occurs as a separate step after loading the array, using 689 nm light tuned close to an excited-state molecular resonance~\cite{Cooper2018,Urech2022}. This collisional loss step could therefore be combined with painted loading to enable the loading of large singe-atom arrays. The model presented here does not include any collisional effects, and we have not seen evidence of a density dependence in our results. 

The largest array we have loaded is the 90-site array spanning 100~\textmu{}m by 200~\textmu{}m shown in figure~\ref{Fig:Pictograph}. However, this is limited by the field of view and (for the number of sites) resolution of our imaging system. A related technique has been used to load an optical lattice $>1$~mm in length~\cite{HobsonThesis}, and we anticipate a similar vertical spatial extent is achievable here. For a loading region of size $200$~\textmu{}m by $1000$~\textmu{}m, a corresponding 2D square array with site spacing of $2$~\textmu{}m would consist of 50 000 sites, increasing up to 5 million if extended to 3D. As the detuning is increased, the horizontal extent of the nMOT also increases, as the resonant shell moves further from the field zero. For reasonable nMOT parameters, a horizontal extent of several hundred micrometers is achievable. Even larger volumes could potentially be loaded by moving the quadrupole magnetic field in 3D during the sequence using computer-controlled shim fields.

\section{Conclusions and Outlook}
\label{sec:Outlook}
In this work, we have introduced painted loading and shown how this method gives full control over the atom number distribution in large Sr tweezer arrays. We have demonstrated the loading of uniform arrays, arrays with a linearly varying atom number and arrays where some sites are not loaded. We have characterised the effects of the lifetime of the atoms in the tweezers (while the nMOT light is still on), the lifetime of the nMOT, the temperature of the nMOT and the rate the nMOT is heated at during the frequency sweep on the number of atoms loaded into sites across the array. By using experimentally measured values for these parameters we were able to construct a rate equation model which reproduced the trends observed in the experimental data.

In future work this technique could be expanded to arrays of single atoms, and from 2D arrays to 3D arrays, where an even larger gain in the number of sites is possible. By combining our model with optimisation tools such as machine learning, we anticipate that significant improvements in array size and loading speed are possible, along with tighter control of the atom number at each site. 

Finally we note that, although demonstrated here for Sr, we expect this technique to be readily extendable to other species which make use of an nMOT, such as Yb, Dy, and Er. With the addition of magnetic field control via shim fields, the technique may be generalisable to any experiment loading arrays from a MOT.

\ack

The work was supported by the Engineering and Physical Sciences Research Council [EP/R035482/1] and by the EMPIR programme co-financed by the Participating States and from the European Union's Horizon 2020 research and innovation programme [23FUN02 CoCoRICO]. We thank the group of Antoine Browaeys for providing the code used to generate our SLM diffraction masks. We also thank Alex Guttridge for his insightful feedback on this work.

\section*{References}
\bibliographystyle{iopart-num}
\bibliography{MainBody.bib}
%\References\bibliography{MainBody.bib}% Produces the bibliography via BibTeX.
\end{document}